\documentclass[aps,pre,twocolumn,superscriptaddress,showpacs]{revtex4}

\usepackage[dvips]{graphicx}
\usepackage{amsmath}

\begin{document}


\title{Bootstrap Percolation on Complex Networks}


\author{G. J. Baxter}
\email[]{gjbaxter@ua.pt}
\affiliation{Departamento de F\'isica, I3N, Universidade de Aveiro,
Campus Universit\'ario de Santiago, 3810-193 Aveiro, Portugal}

\author{S. N. Dorogovtsev}
\author{A. V. Goltsev}
\affiliation{Departamento de F\'isica, I3N, Universidade de Aveiro,
Campus Universit\'ario de Santiago, 3810-193 Aveiro, Portugal}
\affiliation{A. F. Ioffe Physico-Technical Institute, 194021
  St. Petersburg, Russia}
\author{J. F. F. Mendes}
\affiliation{Departamento de F\'isica, I3N, Universidade de Aveiro,
Campus Universit\'ario de Santiago, 3810-193 Aveiro, Portugal}



\date{\today}

\begin{abstract}
We consider bootstrap percolation on uncorrelated complex networks. We
obtain the phase diagram for this process with respect to two
parameters: $f$, the fraction of vertices initially activated, and
$p$, the fraction of undamaged vertices in the graph. We observe two
transitions: the giant active component appears continuously at a
first threshold. There may also be a second, discontinuous, hybrid 
transition at a higher threshold. Avalanches of activations increase
in size as this second critical point is approached, finally diverging at
this threshold. We describe the 
existence of a special critical point at which this second transition
first appears.
In networks with degree distributions whose second moment diverges
(but whose first moment does not), we
find a qualitatively different behavior. In this case the giant active
component appears for any $f>0$ and $p>0$, and the discontinuous
transition is absent. This means that the giant active component is
robust to damage, and also is very easily activated. We also formulate
a generalized bootstrap process in which each vertex can have an
arbitrary threshold.
\end{abstract}


\maketitle
       
\section{Introduction\label{introduction}}

Bootstrap percolation serves as a useful model to
describe in detail or in analogy a growing list of complex phenomena,
including neuronal activity \cite{Eckmann2007,Soriano2008, gdam09},
jamming and rigidity transitions and glassy dynamics
\cite{Sellitto2005,Toninelli2006}, and magnetic systems
\cite{Sabhapandit02}.
Chalupa \emph{et al.} \cite{Chalupa1979} introduced bootstrap percolation in a
particular cellular automaton used to study some magnetic systems (for
other applications see Ref. \cite{Sellitto2005}), see also the even earlier
work of Pollak and Riess \cite{Pollak1975}.
 The standard bootstrap percolation process on a lattice operates as follows:
sites are either active or inactive. Each site is initially active
with a given probability $f$. Sites become active if $k$ nearest
neighbors are active (with $k = 2, 3, ...$). In the final state of
the process, the fraction $\mathcal{S}_{\text{a}}(f)$ of all sites are
active. Remarkably, the function $\mathcal{S}_{\text{a}}(f)$ may be
discontinuous. It may have a jump at a bootstrap percolation threshold
$f_{\text{c2}}$. We will see below that when this process takes place
on a network, this is not the only threshold in this system.

Bootstrap percolation has been thoroughly studied on two and three
dimensional lattices (see \cite{Holroyd2003, Holroyd2006, Balogh2006,
  Cerf1999} and references therein).
The existence of a sharp metastability threshold for bootstrap
percolation in two-dimensional lattices was proved by Holroyd
\cite{Holroyd2003}, and later generalized to $d$-dimensional lattices
\cite{Holroyd2006, Balogh2006}.
More recently, bootstrap percolation has been studied on the random
regular graph \cite{Balogh2007,Fontes2008}, and also on infinite trees
\cite{BaloghPeres06}.  Finite random graphs have also been studied
\cite{Whitney2009}. 
Watts proposed a model of opinions in social networks in which the
thresholds at each vertex is a certain fraction of the neighbors,
rather than an absolute number \cite{Watts2002}.
Bootstrap percolation is closely related to
another well known problem in graph theory, that of the $k$-core of
random graphs \cite{Bollobas84, Pittel96, Fernholz04,
  Dorogovtsev2006a}. 
The $k$-core of a graph is the maximal subgraph for which 
all vertices have at least $k$ neighbors within the
$k$-core. 
It is important to note
the difference between the stationary state of bootstrap percolation
and the $k$-core. Bootstrap percolation is an activation process which
starts from a subset of seed vertices and spreads over a network
according to the activation rules described above. The $k$-core of the
network can be found as an asymptotic structure obtained by a
subsequent pruning of vertices which have less than $k$ neighbors.
While the $k$-core
has been extensively studied, there are no analytical investigations
of bootstrap percolation on complex networks.

In this paper we describe bootstrap percolation on an arbitrary
sparse, undirected, uncorrelated complex network of infinite
size. Specifically, we use the configuration model (a random graph
with a given degree sequence). We show that there are two types of
critical phenomena: a continuous 
transition corresponding to the
appearance of the giant active component, and a second, discontinuous,
hybrid phase transition combining a jump and a singularity. (This
transition is also often called ``mixed''.)
We show that network inhomogeneity strongly influences the critical
behavior at the appearance of the giant active component in networks
with divergent second and third moments and finite first moment of the
degree distribution.
In contrast, the hybrid phase transition has the same critical
singularities for any network with finite second moment of the degree
distribution.
This second
transition can be understood by considering the ``subcritical'' clusters of
the network, consisting of vertices whose number of active neighbors
is one less than the threshold. We show that these subcritical
clusters give rise to avalanches of activations which become
increasingly large as the threshold is approached.
We also describe how the behavior changes when the network is
damaged. The damaging here is the uniformly random removal of
vertices, so that a fraction $p$ of vertices in a network are
retained.  We give the phase diagram showing the thresholds with
respect to both the extent of damage to the network, and to the size
of the initial seed group.
In particular, there is a special critical threshold $p_{s}$ at which
the discontinuous transition first appears. 
We also show that network topology can have a
dramatic effect, as on so called scale-free networks with finite mean
but divergent
second moment of the degree distribution a qualitatively different
behavior occurs. There is no phase transition in the $p$--$f$ plane,
but instead a giant active component appears at any $f\,{>}\,0$ and is
robust to any amount of damage ($p>0$).
Finally, we generalize bootstrap percolation by considering a distribution
of threshold values, so that each vertex may have its own threshold
value. We briefly outline the equations for the active fraction of the
network and the size of the giant active component in this general
formulation, and show how the classical percolation problem and the
usual bootstrap percolation (which we analyze in the remainder of this
paper) are limiting cases.

\section{Results}

Consider an arbitrary, sparse, uncorrelated complex random network in
the infinite
size limit. The structure of this network is completely determined by
its degree distribution $P(q)$. An important (and convenient for
analytical treatment) feature of this architecture is local
tree-likeness, which means that finite loops can be neglected. 
For the time being we assume there are no vertices with degree
zero. We denote by $\langle q\rangle$ the mean of the degree
distribution and similarly $\langle q^2\rangle$ is the second
moment. The network may also be damaged by the uniformly random
removal of vertices so that a fraction $p$ of all original vertices
remaining.

Vertices have either an ``active'' or an ``inactive'' state. Once
activated, a vertex remains active. With probability $f$ each vertex is
part of the seed group, and is in an active state from the start. The
remaining vertices, (a fraction $1-f$) become active only if they have at
least $k$ active neighbors. We iteratively activate vertices that
meet this criterion until a steady state is reached. 

\begin{center}
\begin{figure}[htb]
\includegraphics[width=0.48\textwidth]{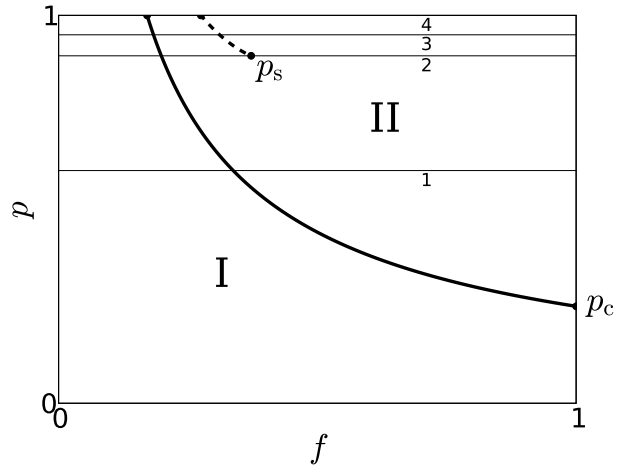}
\caption{Phase diagram of bootstrap percolation in the $f$--$p$
  plane for networks with finite second moment of the degree
  distribution, for $k\geq 2$ and smaller than an upper limit
  $k_{\text{max}}$ determined by the degree distribution.
 The solid line marks $f_{c1}$, the continuous
  appearance of the giant active component from $0$. The giant
  component of active vertices is present above this line in the
  upper-right portion of the diagram (labelled II) and absent in the
  area to the lower-left (I). The dashed heavy curve represents the
  discontinuous transition, $f_{c2}$. This line ends at the special
  critical point $p_{s}$. Thin horizontal lines show the location of
  the traces in Fig.~\ref{traces} relative to the phase diagram
  features. 
}\label{phase}
\end{figure}
\end{center}

We define $\mathcal{S}_{\text{a}}$ to be the fraction of the vertices
in the graph
which are active at equilibrium (that is, including all active
vertices, even those forming finite clusters), which is also the
probability that an arbitrarily selected vertex is active in the final
state of the bootstrap percolation process, and the size of the giant
active component to be $\mathcal{S}_{\text{gc}}$, equal to the
probability that an arbitrarily selected vertex belongs to the giant
active component.  By giant active component we mean a subgraph of
active vertices which forms a connected component that occupies a
finite fraction of the network.

In Fig.~\ref{phase} we show a representative phase diagram for the
giant active 
component in the $f$--$p$ plane for an uncorrelated infinite complex
network whose degree distribution has finite second and third moments.
Results are qualitatively the same for any such network.
The giant active component is absent in the region
labelled I, and present in the region labelled II.
 We see that if the network is sufficiently damaged so that the
 proportion of remaining vertices is less than a critical threshold
 $p_{\text{c}}$, the giant active component never appears, for any
 number of seed vertices. Above $p_{\text{c}} = \langle
 q\rangle/[\langle q^2\rangle-\langle q\rangle]$, which is equal to
 the well known percolation threshold (see, for example
 \cite{dgm2008}),the giant active 
 component appears at some value of $f$, $f_{\text{c1}}$, for a given
 value of $p$. This threshold is marked by the solid heavy line in the
figure. This threshold is above zero for all $p > p_{\text{c}}$.
In the limit $k\to \infty$ the boundary between regions I and II
    tends to the line $pf = p_c$, as the seed vertices may form a
    giant component in the graph. 
For large values of $p$, above a special critical
point $p_{\text{s}}$, 
 we discover a second transition in the size of
the giant active component.
For a given $p > p_{\text{s}}$ there is a threshold $f_{\text{c2}} >
f_{\text{c1}}$ at which the size of the giant active component (and
also the active fraction) jumps suddenly.
These points are marked by the heavy dashed line in Fig.~\ref{phase}.
Note that Fontes and Schonmann \cite{Fontes2008} noticed on undamaged
regular random graphs the two transitions we observe in more complex
networks.

\begin{center}
\begin{figure}[htb]
\includegraphics[width=0.48\textwidth]{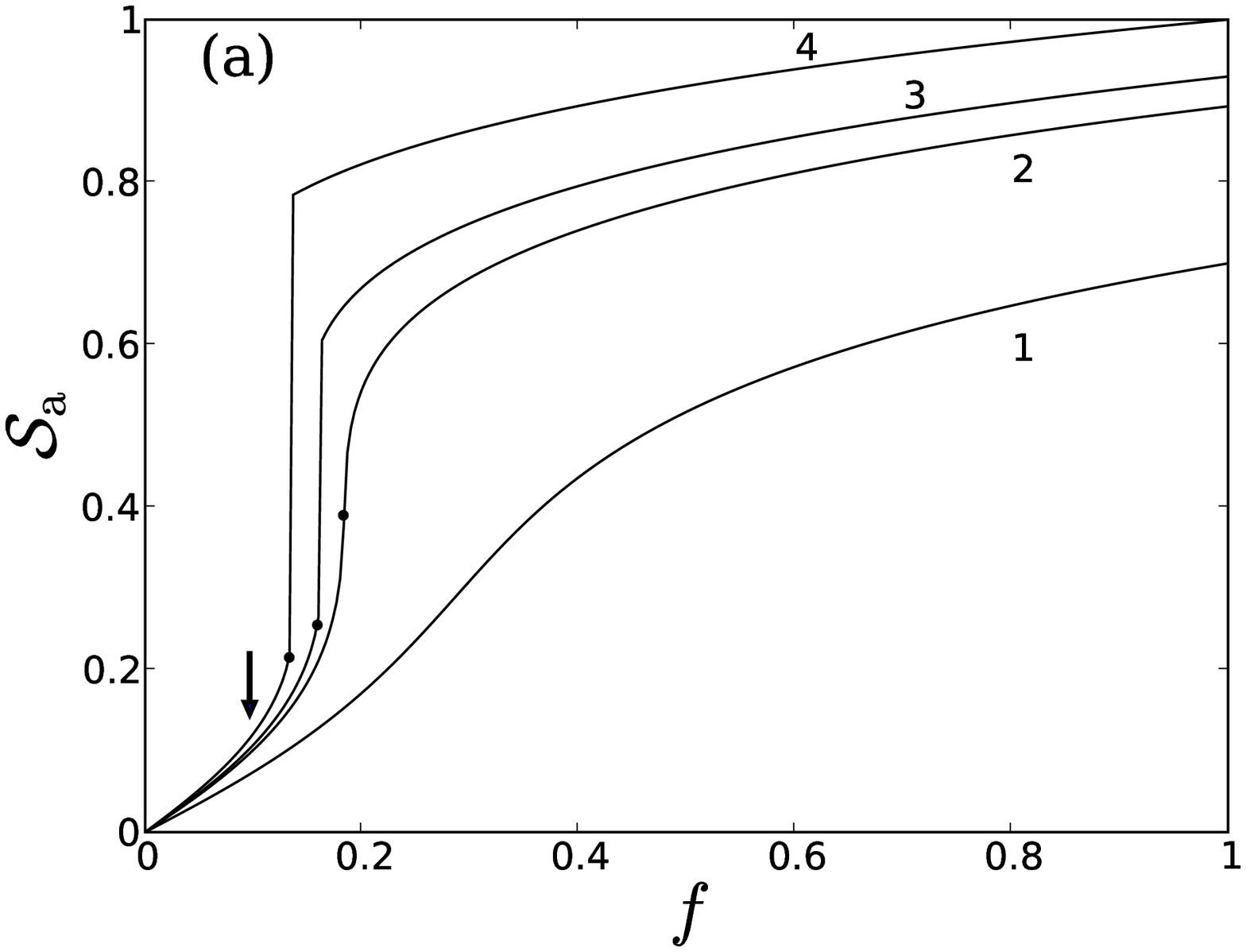}
\includegraphics[width=0.48\textwidth]{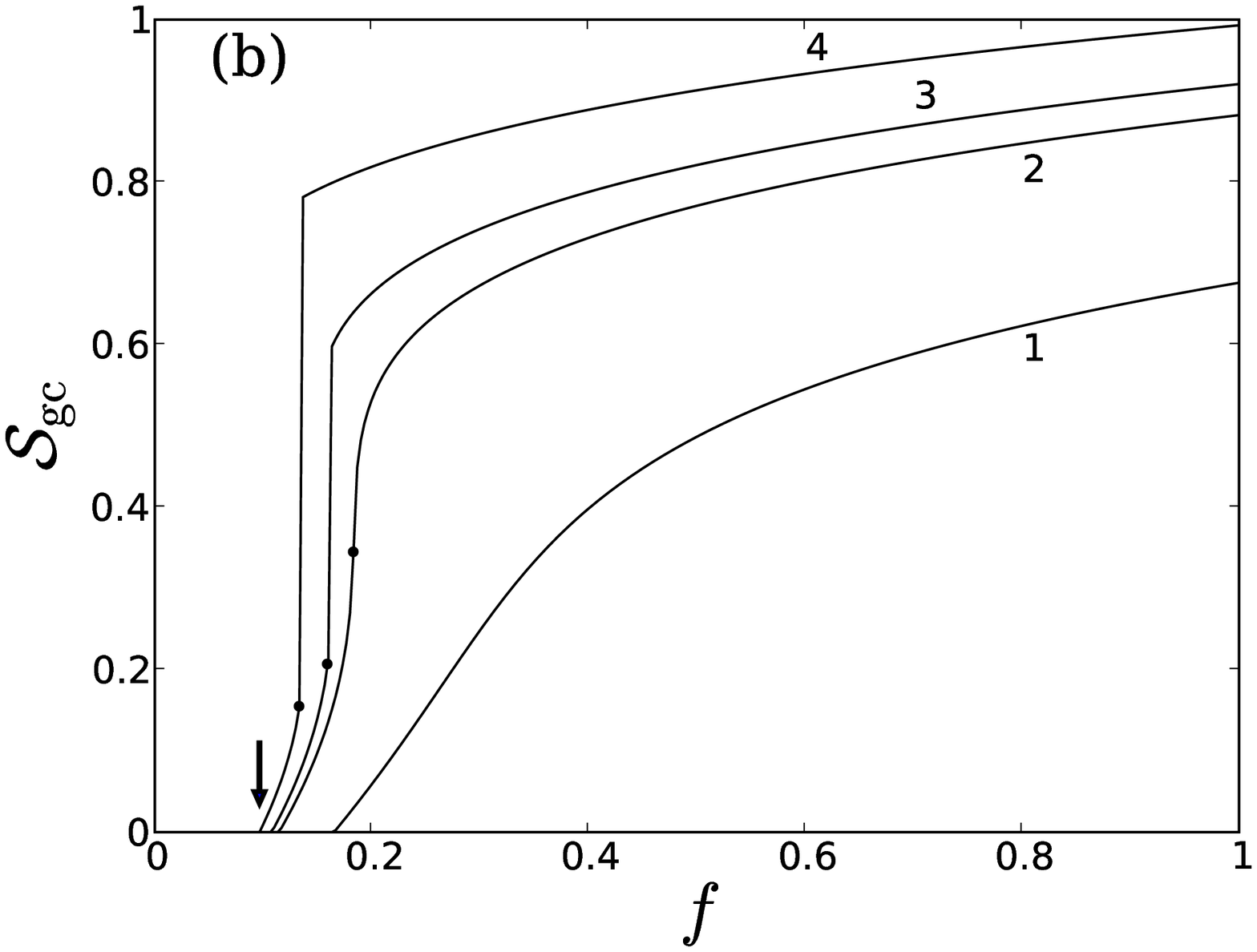}
\caption{Probability that an arbitrarily chosen vertex is (a)
active and (b) in a giant connected component
of active vertices for an Erd\H{o}s-R\'{e}nyi graph of mean degree
$5$, with $k=3$. The four lines (labelled from 1 to 4) are: $p=\{0.7,
0.893, 0.93,1\}$,
corresponding to the relative positions in the phase diagram shown as
thin horizontal lines in Fig.~\ref{phase}, namely: 1. $p_{\text{c}} < p
< p_{\text{s}}$; 2. $p = p_{\text{s}}$; 3. $p_{\text{s}} < p < 1$ and
4. $p =1$. The arrows in each plot mark the point of emergence
($f_{\text{c1}}$) of 
the giant active component for $p=1$. The small dot on each trace
marks the point of the hybrid transition.}\label{traces} 
\end{figure}
\end{center}

Fig.~\ref{traces} shows the active fraction $\mathcal{S}_{\text{a}}$
and the
size of the giant active component $\mathcal{S}_{\text{gc}}$ as a
function of $f$ for four values of $p$ in an Erd\H{o}s-R\'{e}nyi graph
(which has a Poisson degree distributions in the infinite size
limit). We choose this network as it is representative of random
graphs. For comparison, the position of each in the phase diagram is
marked by a faint solid line in Fig.~\ref{phase}. Line $1$ is for a
value of $p$ before the appearance of the jump. Line $2$ is exactly at
the special critical point $p_{\text{s}}$ at which the jump appears.
 Line $3$ is at a $p > p_{\text{s}}$ where there is a jump.
The location of the jump moves to smaller values of $f$ as $p$
increases, but never reaches zero, as is demonstrated by line $4$,
which is at $p=1$.
The giant active component
 appears continuously and linearly from zero, exactly as it does in
 ordinary percolation \cite{AlbertBarabasi02,Dorogovtsev2002,dgm2008}.
It is interesting to note that there is no discontinuity in
$\mathcal{S}_{\text{a}}$ at $f_{\text{c1}}$ (marked by a small arrow),
i.e., the threshold is invisible (hidden) when observing only the
overall activation of the network.

For $k=1$ the jump does not appear, we have only the continuous
transition. For larger $k$ there is a jump, and it appears
at larger values of $f$ (for a given $p$) the larger $k$ is. 
The value
of $p_{s}$ also increases, such that there is a finite maximum
threshold, $k_{\text{max}}$ (proportional to the mean degree of the
network for Erd\H{o}s-R\'{e}nyi graphs) beyond which the jump no
longer appears.  
That is, the dashed line in Fig.~\ref{phase}, which marks the location
of the jump, moves to the right and towards the top of the graph as
$k$ increases, finally disappearing completely above $k_{\text{max}}$.
This discontinuous transition has a hybrid character, with both a
discontinuity and 
a singularity: when approaching $f_{\text{c2}}$ from below the size of
the giant active component approaches the value at the bottom of the
jump as the square-root of the distance from $f_{\text{c2}}$:
\begin{equation}\label{scaling1}
\mathcal{S}_{\text{a}}(f) = \mathcal{S}_{\text{a}}(f_{\text{c2}}) -
a(f_{\text{c2}}-f)^{1/2},
\end{equation}
where $a$ is a constant (see Section~\ref{analysis} below for the
origin of this equation). Lines $3$ and $4$ in Fig.~\ref{traces}
illustrate this situation.
The same result holds with respect to $p$
if we were to approach this jump along a line of constant $f$.
The size of the giant active component, $\mathcal{S}_{\text{gc}}$, has
the same critical behavior.

The height of the jump decreases with decreasing $p$ (while
$f_{\text{c2}}$ increases slightly), disappearing at the special point
$p_{\text{s}}$. (The line labelled $2$ in Fig.~\ref{traces} is at $p =
p_{\text{s}}$.) At this point the behavior is different, as the size
of the giant component approaches $f_{\text{c2}}$ now as the cube-root
of the distance from the threshold (see Section~\ref{analysis}):
\begin{equation}\label{scaling2}
\mathcal{S}_{\text{a}} = \mathcal{S}_{\text{a}}(f_{\text{c2}}) -
a'(f_{\text{c2}}-f)^{1/3},
\end{equation}
where $a'$ is a constant.

\begin{center}
\begin{figure}[htb]
\includegraphics[width=0.23\textwidth,angle=0]{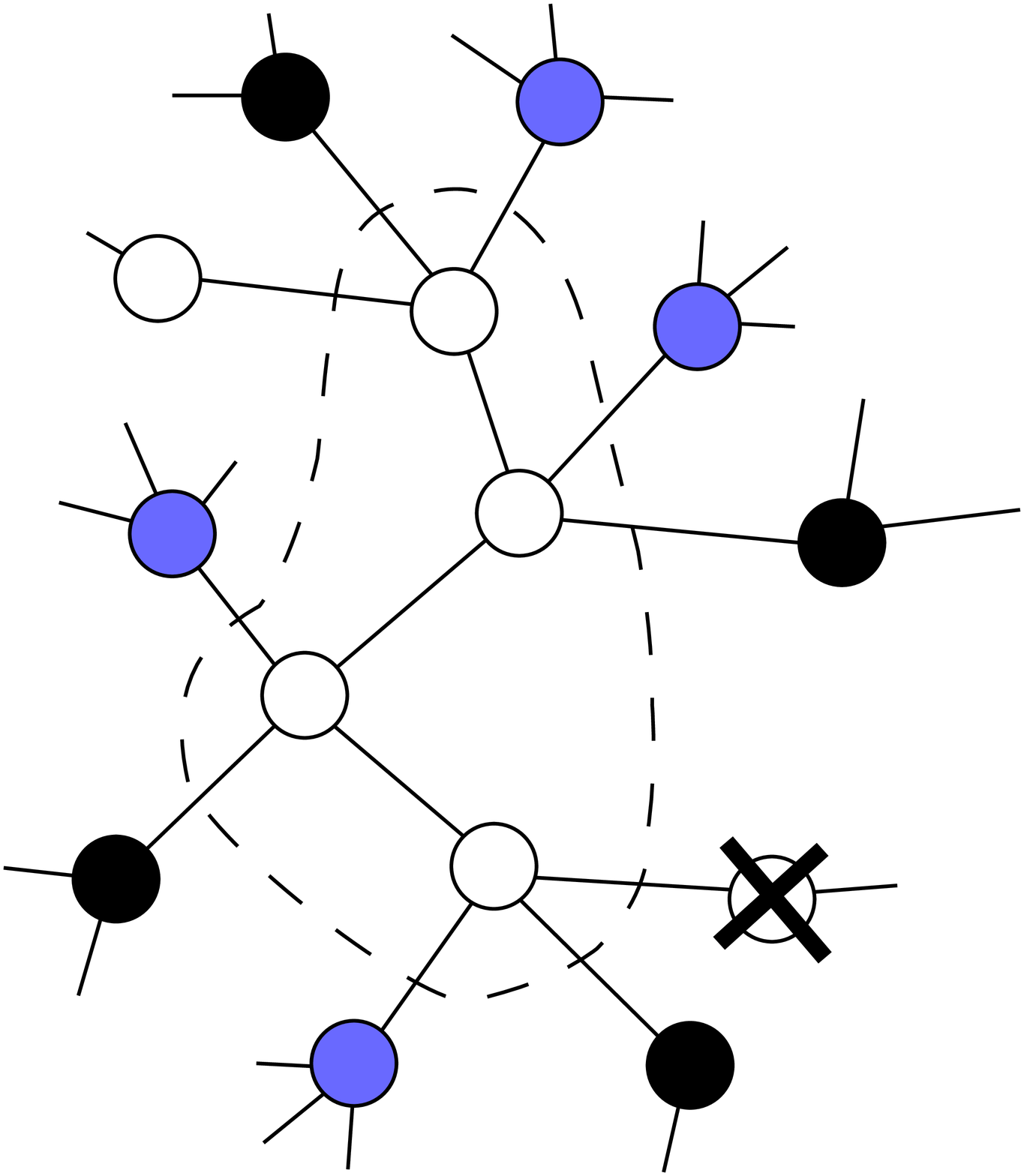}\hfill
\includegraphics[width=0.23\textwidth,angle=0]{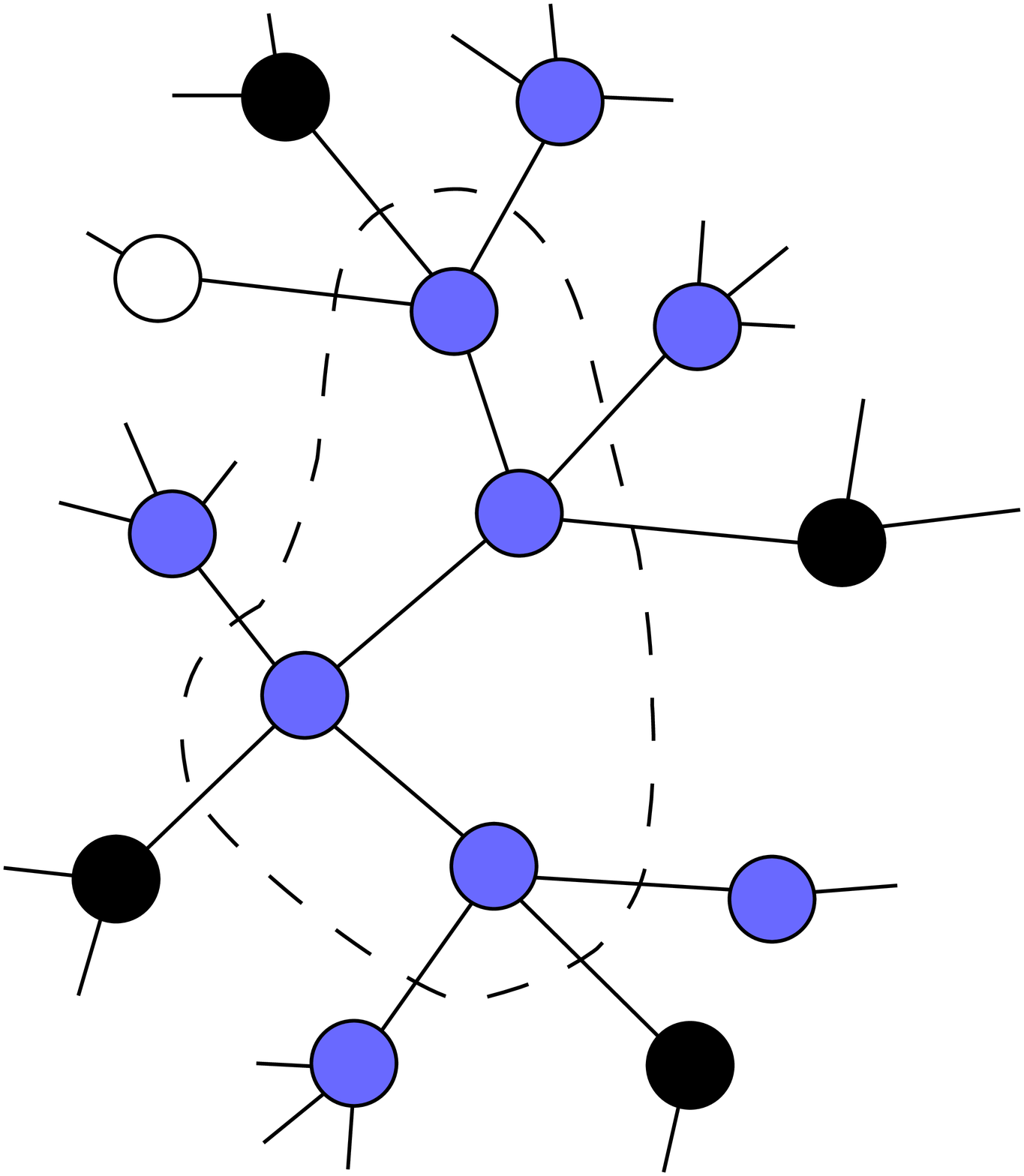}
\caption{Left: A subcritical cluster in a network with threshold $k = 3$. Filled
  black vertices are seed vertices, shaded vertices are active
  vertices while empty vertices are inactive. 
The vertices inside the dashed loop form a subcritical cluster, while those connected to them are either
inactive or have more than the threshold number of active
neighbors. If a single vertex neighboring a subcritical cluster becomes
  active (for example, if it became a seed vertex)---here marked by a cross
 ---it's neighbor inside 
  the subcritical cluster must become
  active, and then it's neighbors, until the entire cluster
  is activated, as shown on the right.
}\label{corona-1}\label{corona}   
\end{figure}
\end{center}

To understand the discontinuous ``jump'' in the size of the active
component of the graph, we consider the subcritical clusters, a
concept related to the corona clusters which were used to describe a
similar transition in
$k$-core percolation \cite{Goltsev2006,Dorogovtsev2006b,
  Schwarz2006}. The subcritical vertices of the graph are the vertices
whose number of active neighbors is precisely one less than the 
threshold of activation for that vertex.
An example of
a small subcritical cluster is illustrated in Fig.~\ref{corona}.

Clusters of subcritical vertices are important because of the
following quality.
 The activation of even a
single vertex 
neighboring the subcritical cluster necessarily leads to at least one
of the members of the cluster now meeting its activation threshold. In
turn, this will activate one of its neighbors in the cluster, and so
on, so that an avalanche of activations ensues until the entire
subcritical cluster becomes active---see Fig.~\ref{corona-1}.
 Below and above the jump, the subcritical vertices form only finite
 and isolated clusters, but as $f_{\text{c2}}$ is approached from 
below, the mean size of the subcritical clusters diverges.
Hence the
avalanches resulting from the change in activation state of a single
vertex form a finite fraction of the entire graph, leading to
a discontinuous change in the size of the active fraction of the
graph. This argument will be made more precisely in the analysis in
Section~\ref{avalanches} below.

When the degree distribution of the network decays very slowly,
specifically in networks with divergent second moment $\langle q^2
\rangle$ or third moment $\langle q^3 \rangle$ of the degree
distribution, the results are different from those described above. In
particular, this is the case if the
degree distribution tends to the form $P(q) \propto q^{-\gamma}$ with
$\gamma \leq 4$ for large $q$. If $\gamma \leq 2$ the mean of the
degree distribution also diverges, but we do not consider this case
here.
If $2 < \gamma \leq 3$, the second moment diverges.
These scale-free networks are of particular interest
because many large real natural and technological networks appear to
be of this kind \cite{AlbertBarabasi02,Dorogovtsev2002}. 
In this case (in the limit $N\to\infty$) a finite fraction of vertices is
activated at any $p$ or $f > 0$, and for any arbitrary activation
threshold $k$.
In other words the location of the jump tends
to zero as the size of the network increases---so in very large
scale-free networks we will not find (at finite $f$ or $p$) either of
the transitions observed in graphs with fast decaying degree distributions.
This means that, as has
been found in several other cases \cite{Albert2000,Cohen2000}, such
scale-free networks are very robust to damage, and also that
such a network is very easily activated.

When $3 < \gamma \leq 4$, the phase diagram is qualitatively the same
as that shown in Fig.~\ref{phase}. The giant active component appears
at finite $p$ ( or $f$) with a continuous transition. However, rather
than growing linearly near the transition point, the size of the giant
active component increases as the distance from the critical point
raised to the power $1/(\gamma-3)$.

\section{Basic Analysis\label{analysis}}

In this section and the two following, we describe in more detail how
the results already described may be obtained. 

Consider choosing an arbitrary vertex from the network. We wish to
calculate the probability $\mathcal{S}_{\text{a}}$ that this vertex is
active in the equilibrium state.  
To calculate this probability, we first define $Z$ as follows:
$Z$ is the probability
that, on following an arbitrary edge in the graph, we reach a vertex 
which is either a seed vertex or has at least $k$ downstream 
neighbors that are active. (By downstream we mean neighbors of the
vertex reached by the edges other than the one we arrived from.)
To be active, these downstream neighbors in turn must fulfil this same
condition, that they are either seed vertices or they have $k$ further
downstream
neighbors of their own that are previously active. 

\begin{table}[htb]
\includegraphics[width=0.25\textwidth]{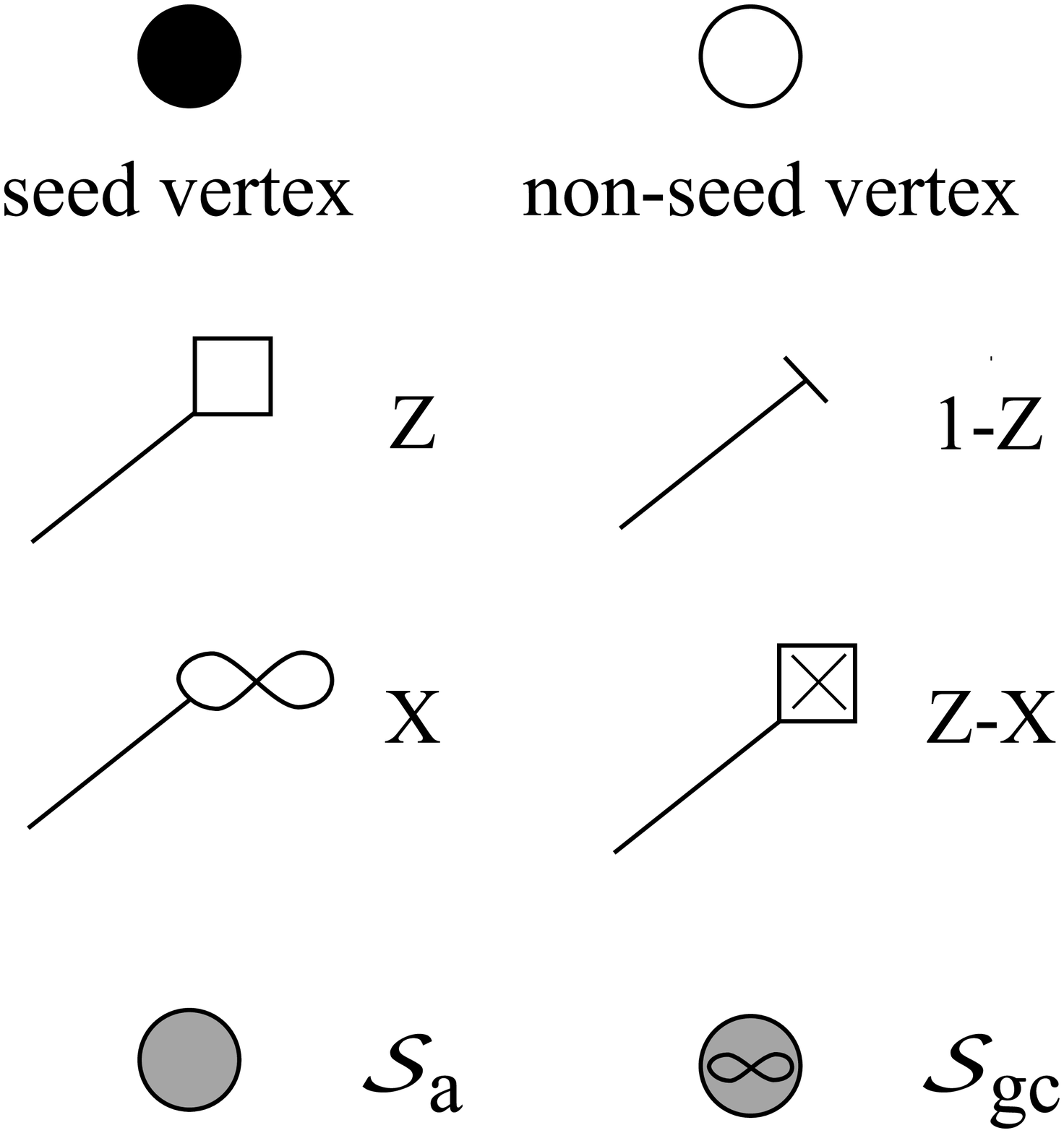}
\caption{Symbols used in graphical representations of self-consistency
  equation.\label{symbols}}
\end{table}

We can graphically represent this recursive relationship using the
symbols given in Table \ref{symbols}. The probability $Z$ is
represented by an edge ending in a square. A seed vertex is represented
by a black disc, and other vertices by open discs. An edge crossed by
a short line at its end represents the probability $1-Z$, that is the
probability of encountering a vertex that doesn't satisfy the
condition for $Z$. Thus we obtain the following representation for $Z$:
\begin{center}
\includegraphics[width=0.35\textwidth]{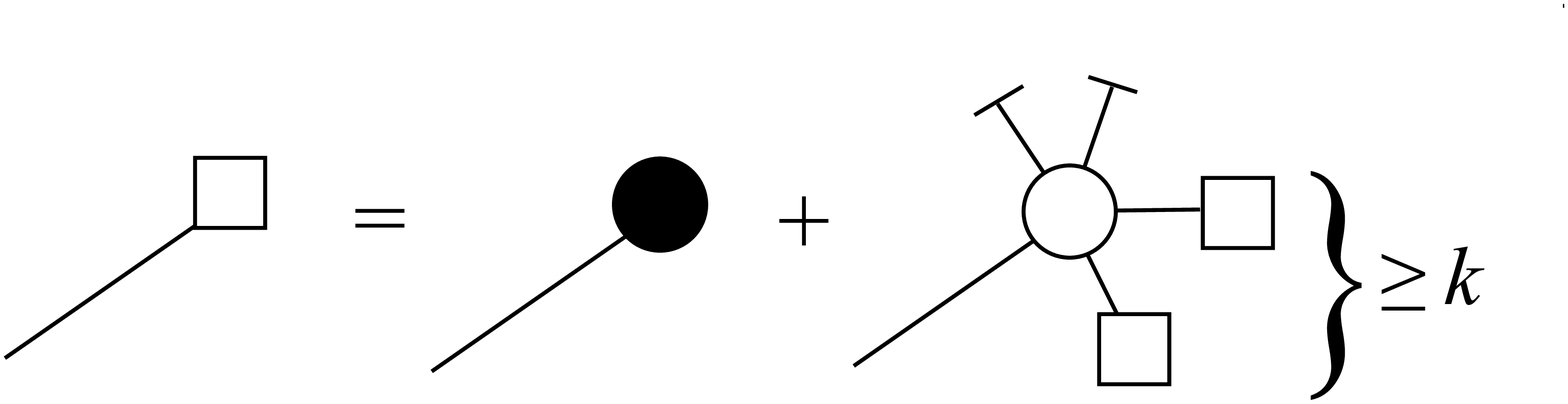}
\end{center}
The
terms on the right hand side represent sums of the probabilities of
all such terms.
Based on this diagram, we write mathematical expressions for the
probabilities represented by each of these symbols, allowing us to construct
the following  self-consistency equation for $Z$:
\begin{align}
Z  = pf + p(1-f) \sum_{i =  k}^{\infty} &
\frac{(i+1)P(i+1)}{\langle q\rangle }
\nonumber
\\[5pt]
&\times\sum_{l=k}^{i} \binom{i}{l}Z^l(1-Z)^{i-l}.
\label{Zfull}
\end{align}
The probability $\mathcal{S}_a$, represented by a shaded disc, is the
sum of two terms, as represented in this diagram:
\begin{center}
\includegraphics[width=0.28\textwidth]{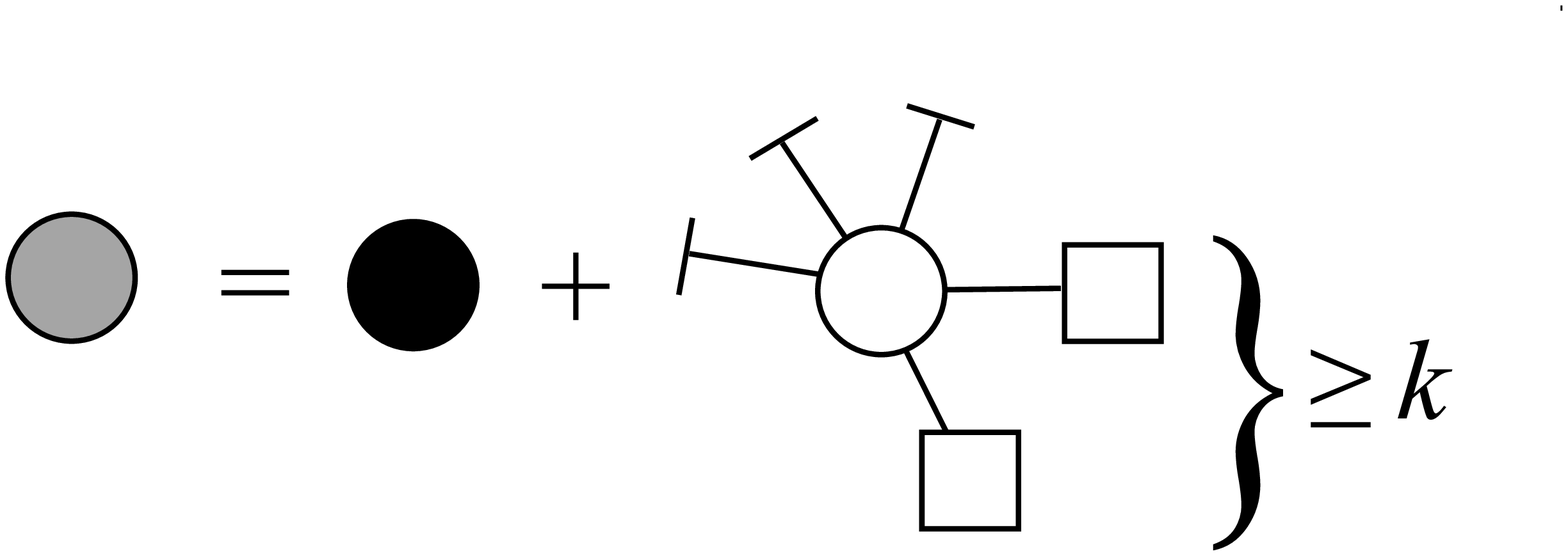}
\end{center}
The first is the probability that the vertex is active from the
beginning ($pf$), the second [with prefactor $p(1-f)$] is the
probability that it has at least $k$ neighbors that would be active
even if the vertex we are observing was inactive. But each neighbor
satisfies this condition precisely with probability $Z$,  as
represented by squares in the diagram. 
Converting to a mathematical
expression, this gives the following equation:
\begin{align}
&\mathcal{S}_{\text{a}} = pf + p(1-f)\sum_{i =  k}^{\infty} 
P(i)\sum_{l=k}^{i} \binom{i}{l}Z^l(1-Z)^{i-l}.
\label{Sa_full} 
\end{align}

The probability $\mathcal{S}_{\text{gc}}$ that an arbitrarily chosen vertex
belongs to the giant active component can be constructed in a similar
way, but we must impose the further condition that a vertex has an edge
leading to an active subtree of infinite extent. We define $X$ to be the
probability that the vertex encountered upon following an arbitrarily
chosen edge meets the conditions for $Z$ and also has an edge
leading to an active subtree of infinite extent. 
Graphically, we represent the probability $X$ by an infinity symbol at
the end of an edge, and a self-consistency condition for $X$ is
expressed by the diagram:
\begin{center}
\includegraphics[width=0.40\textwidth]{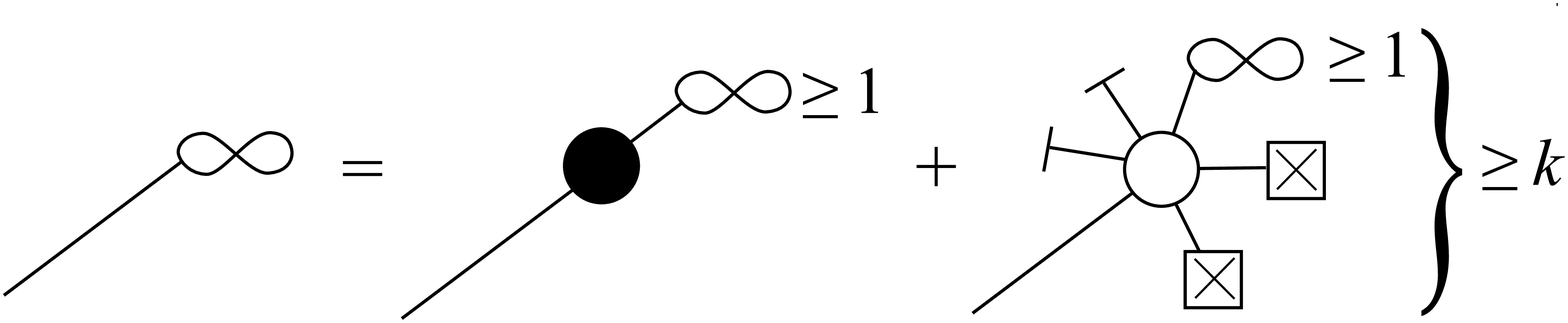}
\end{center}
This corresponds to the equation:
\begin{align}
X = pf & \sum_{i=0}^{\infty} \frac{(i+1)P(i+1)}{\langle q\rangle }
\sum_{m=1}^i\binom{i}{m}X^m(1-X)^{i-m}
\nonumber\\[5pt]
 +& p(1-f) \sum_{i=k}^{\infty}\frac{(i+1)P(i+1)}{\langle q\rangle }
\sum_{l=k}^i\binom{i}{l} 
\nonumber\\[5pt]
&{\times}
 \sum_{m=1}^l\binom{l}{m}X^m(Z-X)^{l-m}(1-Z)^{i-l}.
\label{Xfull}
\end{align}
The probability $\mathcal{S}_{gc}$ (the probability that an
arbitrary vertex belongs to a giant active component, represented below
by a shaded circle containing the infinity symbol---see
Table \ref{symbols}) is the sum of the
probability that the vertex is a seed vertex that is connected to an
infinite active subtree (probability $X$) and of the probability that
the vertex is not a seed vertex but has at least $k$ independently active
neighbors (probability $Z$), at
least one of which also leads to an infinite active subtree. Thus
\begin{center}
\includegraphics[width=0.36\textwidth]{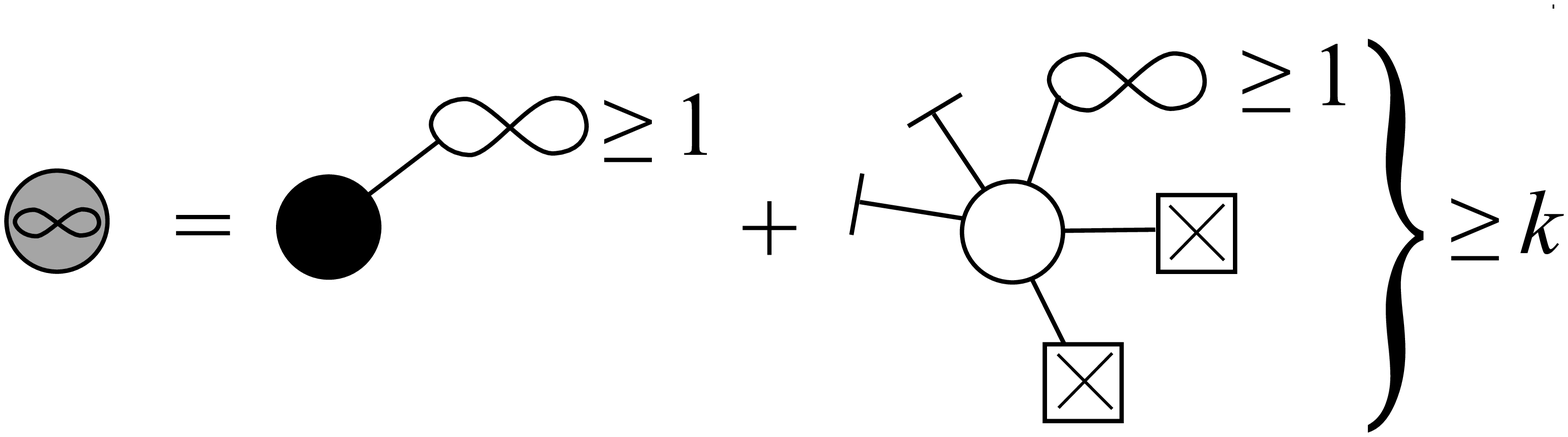}
\end{center}
so that, similarly to Eq. (\ref{Sa_full}), we can write
$\mathcal{S}_{\text{gc}} $ in terms of $X$ and $Z$:
\begin{align}
\mathcal{S}_{\text{gc}} =& pf \sum_{i=0}^{\infty} P(i)
\sum_{m=1}^iX^m(1-X)^{i-m}
\nonumber\\[5pt]
& +  p(1-f) \sum_{i=k}^{\infty}P(i) 
\sum_{l=k}^{i}\binom{i}{l}
\nonumber\\[5pt]
&\times \sum_{m=1}^{l}\binom{l}{m}
X^m(Z-X)^{l-m}(1-Z)^{i-l} 
.
\label{Sgc_full}
\end{align}
It will be useful to define $\Psi(Z,p,f)$ to be the right-hand side of Eq.~(\ref{Zfull}), 
and $\Phi(X,Z,p,f)$ the right-hand side of Eq.~(\ref{Xfull}), so that these
two equations become
\begin{equation}
\label{Zpsi}
Z = \Psi(Z,p,f),
\end{equation}
and
\begin{equation}
\label{Xphi}
X = \Phi(X,Z,p,f)
.
\end{equation}

These equations can be solved numerically for a given network degree
distribution.  If multiple solutions exist, the physical solution for
$Z$ is always the smallest value.
The location, $f_{c1}$ of the appearance of the giant active component can be
found by assuming $X$ is small but non-zero in Eq.~(\ref{Xphi}),
taking the limit as $X$ tends to zero and solving for $f$ for a given
$p$ (or vice versa). In this way we also find that $X$ and hence
$\mathcal{S}_{\text{gc}}$ grow linearly from the critical point
$f_{c1}$ for networks with $\langle q^2 \rangle$ finite. The results
mentioned below also correspond to this case---we will examine the case
 $\langle q^2 \rangle \to \infty$ subsequently.

The second, discontinuous, transition can be located by observing that
the jump occurs when, (after a second solution appears) the smallest
$Z$ solution of Eq.~(\ref{Zpsi}) disappears. At this point $\Psi(Z)$ just
coincides with the value of $Z$,
and a little consideration reveals that this must be at a local
extremum of $\Psi/Z$. Thus the discontinuous transition can be found
by simultaneously solving Eq.~(\ref{Zpsi}) and
\begin{equation}\label{Zjump}
\frac{d}{dZ}\left(\frac{\Psi}{Z} \right) = 0
\end{equation}
for $f$.
The fact that the first derivative vanishes leads to the square-root
scaling near the critical point, with respect to either $f$ or
$p$---see Eq. (\ref{scaling1}). 
The jump disappears at a special critical point $p_{\text s}$ in the
$f-p$ plane which satisfies Eqs.  (\ref{Zfull}), (\ref{Zjump}) and a
third condition 
\begin{equation}\label{Zjump_last}
\frac{d^2}{dZ^2}\left(\frac{\Psi}{Z} \right) = 0.
\end{equation}
This condition means that the scaling below $p_s$ (see
Fig.~\ref{phase}) is cube-root---see Eq. (\ref{scaling2}).

\section{Avalanches\label{avalanches}}

The singular behavior [Eq. (\ref{scaling1})] near the hybrid transition 
can be understood by considering the
subcritical clusters of the active subgraph. These form a subset of the
inactive portion of the graph consisting only of those vertices whose
number of active neighbors is exactly one less than the activation threshold
for that vertex---see Fig. \ref{corona}. 
That is, the subcritical subgraph consists of all those vertices
which are not seed vertices and which have
exactly $k-1$ active neighbors.
The subcritical clusters are finite everywhere
except exactly at the point of the jump transition.
To show that this is the case, we use a generating function approach,
similar to that used 
in \cite{Goltsev2006,Callaway2000,Newman2001}, to
calculate $\langle s\rangle_{\text{sub}}$, the mean
size of the subcritical clusters. 

Let $F_0(x)$ be the generating function for the probability that an
arbitrarily chosen vertex is
subcritical. By considering the probability that an arbitrarily chosen
vertex is subcritical, which corresponds to $F_0(1)$, we can write
\begin{align}
F_0(x)=& p(1-f) \nonumber\\[5pt]
&\times\sum_{q\geq
  k-1}P(q)\binom{q}{k-1}Z^{k-1}(1-Z)^{q-k+1}x^{q-k+1}.\label{F0}
\end{align}
Similarly, the generating function for the probability that an
arbitrarily chosen edge leads to a subcritical vertex is
\begin{equation}
F_1(x)= p(1-f) \sum_{q\geq k}\frac{qP(q)}{\langle q\rangle
}\binom{q-1}{k-1}Z^{k-1}(1-Z)^{q-k}x^{q-k}.\label{F1}
\end{equation}

The generating function for the probability that a randomly chosen
vertex belongs to a subcritical cluster of a given size then must obey
the self-consistency equation \cite{Callaway2000,Newman2001}
\begin{equation}
H_0(x) = 1 - F_0(1) + xF_0[H_1(x)],\label{H0}
\end{equation}
where $1 - F_0(1)$ represents the probability that the randomly chosen
vertex is not 
itself subcritical, and the second term is a recursive relationship,
ensuring that successive powers of $x$ correspond to the probabilities
of encountering a cluster size matching that power.
In this equation $H_1(x)$ is the related generating function for the
probability
that a subcritical cluster of a given size is reached upon following
an arbitrarily chosen edge. In a similar way we can write a
self-consistency equation for this:
\begin{equation}
H_1(x) = 1- F_1(1) + xF_1[H_1(x)].\label{H1}
\end{equation}
Where $1-F_1(1)$ is the probability that the arbitrarily chosen edge
leads to a vertex that is not subcritical. Note that $H_0(1) = H_1(1)
= 1$.

From these generating functions, we can calculate various quantities
related to the subcritical clusters. For example, the distribution of
avalanche sizes (which are the same as the sizes of the subcritical
clusters) is given by
\begin{equation}
G(s) = \frac{1}{s!}\frac{d^sH_0(x)}{dx^s}\bigg\vert_{x=0}
\end{equation}
and we expect that at the critical point $G(s)\sim s^{-3/2}$.
The mean size of the subcritical clusters is simply
\begin{align}
\langle s\rangle_{\text{sub}} = &
\frac{dH_0}{dx}\bigg\vert_{x=1}\nonumber\\[5pt]
= & F_0(1) + p(1-f)\sum_{q\geq k-1}P(q)\binom{q}{k-1}
\nonumber\\[5pt]
&\times Z^{k-1}(1-Z)^{q-k+1}(q-k+1)\frac{dH_1}{dx}\bigg\vert_{x=1}
.
\end{align}
Using Eqs. (\ref{H1}), (\ref{F1}) and comparing with Eqs. (\ref{Zfull})
and (\ref{Zpsi})
we find that
\begin{equation}
\frac{dH_1}{dx}\bigg\vert_{x=1} = \frac{F_1(1)}{1 - d\Psi(Z)/dZ}.
\end{equation}
Now from Eq. (\ref{Zjump}), $ d\Psi(Z)/dZ =1$ at the critical point, and
 $1 - d\Psi(Z)/dZ \propto Z - Z_j \propto
(f_{\text{c2}}-f)^{1/2}$,
 near the critical
point. Thus, near this point, therefore, the term containing
$dH_1/dx\big\vert_{x=1}$
 dominates $\langle
s\rangle_{\text{sub}}$ so that
\begin{equation}
\langle s\rangle_{\text{sub}} \propto (f_{\text{c2}}-f)^{-1/2}
,
\end{equation}
or alternatively, for fixed $f$, $\langle s\rangle_{\text{sub}} \propto (p_{\text{c2}}-p)^{-1/2}$, 
hence the mean size of the corona clusters diverges at the critical point.

The addition of a single vertex (an
infinitesimal increase in $p$) or activation of a seed vertex (increment
of $f$) may lead to the activation of a subcritical vertex
and hence 
activating an entire subcritical cluster in an avalanche. 
At $f_{\text{c2}}$, the subcritical clusters span the whole graph, so the
activation of a vertex can lead to an avalanche of activation that
eventually affects a finite fraction of the whole infinite graph
hence we see a discontinuity in both the size of the active fraction
and the giant active component. Note that for $f>f_{c2}$ the mean size
of the subcritical clusters is finite.

\section{Scale-free graphs\label{scalefree}}

Let us consider degree distributions that tend to $P(q) \propto
q^{-\gamma}$ for large $q$ where exponent $\gamma$ is some positive constant
usually $> 2$. For concreteness, in the following we will
consider the degree distribution
\begin{equation}
P(q) = Aq^{-\gamma} \qquad \mbox{ for } q \geq q_0,
\end{equation}
where $A$ is a constant of normalization.
For $\gamma > 4$ the results are qualitatively the
same as those described above. 

When $\gamma \leq 3$ 
the
second moment of the degree distribution diverges, leading to
different behavior. The results that follow refer to the situation
when $2 < \gamma \leq 3$. Note that many real world networks, especially
biological networks have exponent in the range $1 < \gamma \leq 2$
\cite{AlbertBarabasi02,Dorogovtsev2002,Bonifazi2009}. In this case
 the first moment also
diverges. We don't consider this case here.

By assuming $Z$ to be small, we can approximate $\Psi(Z)$ by considering
only leading order in $Z$. Then the self
consistency equation (\ref{Zfull}) becomes:
\begin{equation}\label{Z_sf}
Z \approx p(1-f)aZ^{\gamma-2} + pf.
\end{equation}
For $p>0$, this equation has no small-$Z$ solution, even in the limit
$f\to 0$. Because the only solutions for $Z$ as $f\to 0$ are therefore of
order $1$ it is clear that the discontinuous transition is moved to
$f=0$ for scale-free graphs.
A similar analysis for the giant active component---approximating
$\Phi(Z,X)$ [the RHS of Eq. (\ref{Xfull})] by assuming $X$ and $Z$ both
small leads to similar
conclusion about $X$: that there are no infinitesimal solutions in
the limit $f \to 0$, confirming that there is no jump for $f>0$ but
also that the giant active component appears for any $f>0$ and $p>0$.

To add support to this approximation, consider
the same degree distribution as before, but truncated at some maximum
degree $q_{\text{cut}}$ (the normalization constant will also necessarily
change). If we re-derive (\ref{Z_sf}) assuming a finite $q_{\text{cut}}$, we
find
\begin{equation}\label{Z_sf_qmax}
Z \approx p(1-f)bZ^{k} + pf
\end{equation}
which does have a solution at finite $p$ (or $f$).
For finite $q_{\text{cut}}$, numerical solution of Eq. (\ref{Zfull}) shows a
jump appears at small values of $f$. As $q_{\text{cut}}$ is increased, the
curve of this jump moves closer to $f=0$, and extends towards
$p=0$. Similarly the giant component appears at smaller and smaller
values of $p$ and $f$ as $q_{\text{cut}}$ is increased. In keeping with the
approximate analysis just described, we expect that both thresholds
reach $f=0$ and $p=0$ in the infinite size limit.
In summary, when $2<\gamma \leq 3$, the giant active component is
always present everywhere in 
the $p - f$ plane for $p > 0$ and $f > 0$, and appears not from zero
but with a finite size.

When  $3<\gamma \leq 4$, an expansion of the right-hand-side of
Eq.~(\ref{Xfull}) in leading powers of $X$ gives an equation of the form
\begin{equation}\label{Xgamma4}
X = c_1X + c_2X^{\gamma-2} + ...\,,
\end{equation}
where the ellipsis signifies further terms of higher order in $X$.
The first coefficient $c_1 = pf\frac{\langle q^2\rangle - \langle
  q\rangle}{\langle q\rangle} + p(1-f)B(Z)$, where $B(Z)$ is a
function of the variable $Z$. Thus when $f < 1$
the value of $c_1$ differs from that found in the percolation
problem. 
The presence of a finite linear term ($c_1 > 0$)
means that the appearance of the giant component occurs at non-zero
values of $p$ (or $f$) -- at a point which can be found by solving
$c_1 = 1$. 
 However, because the second
leading exponent is $\gamma-2$ and not $2$, $X$ scales as
$(p-p_{c1})^{\beta}$, near the appearance of the giant
component, with $\beta = 1/(\gamma-3)$. This is the same scaling as
found in the usual percolation problem \cite{Cohen2002}. Curiously, the
second coefficient $c_2$ is simply equal to $p$ up to a 
factor depending on the degree distribution.
Above
$\gamma=4$, we find $\beta = 1$ as found for the usual percolation.

\section{General distribution of activation thresholds $k$}\label{general}

The bootstrap percolation process described above can be thought of as
a specific case of a more general process in which the threshold
values can be different for each vertex. Assuming no correlations
between vertex degree and threshold value, we can define a
distribution $Q(k)$ 
such that $Q(k)$ is the fraction of vertices which have threshold
value $k$. The fraction of seed vertices is then $Q(0)$.
Setting $Q(0) = f$ and $Q(k)= 1-f$ for some $k \geq 2$ we recover the
bootstrap percolation model described above.

In the general case, we find that the equation for the active fraction
is
\begin{align}
\mathcal{S}_{\text{a}} = p
\sum_{k\geq 0} Q(k)
\sum_{i =  k}^{\infty} 
P(i)\left[\sum_{l=k}^{i} \binom{i}{l}Z^l(1-Z)^{i-l}\right],
\label{Sa_gen} 
\end{align}
where, as above, $Z$ is the probability of encountering a vertex with
at least $k$ downstream active neighbors upon following an arbitrary
edge:
\begin{align}
Z  =  p\sum_{k\geq0} Q(k)\sum_{i =  k}^{\infty}
\frac{(i+1)P(i+1)}{\langle q\rangle }\sum_{l=k}^{i} \binom{i}{l}Z^l(1-Z)^{i-l}.
\label{Zgen}
\end{align}
These two equations are similar to those presented in
\cite{Gleeson2008} for undamaged networks as a generalization of the
Watts model \cite{Watts2002}.

Similarly, the equation for the giant active component is
\begin{align}
\mathcal{S}_{\text{gc}} 
=p\sum_{k\geq0} Q(k)& \sum_{i=k}^{\infty}P(i)\sum_{l=k}^{i}\binom{i}{l}\nonumber\\[5pt] 
&\times\sum_{m=1}^{l}\binom{l}{m}X^m(Z-X)^{l-m}(1-Z)^{i-l},
\label{Sgc_gen}
\end{align}
where as before $X$ is the probability that an edge leads to an
infinite active subtree:
\begin{align}
X =& 
p\sum_{k\geq0} Q(k) \sum_{i=k}^{\infty}\frac{(i+1)P(i+1)}{\langle q\rangle }
\nonumber\\[5pt]
&{\times}
\sum_{l=k}^{i} \binom{i}{l}
\sum_{m=1}^{l} \binom{l}{m}
X^m(Z-X)^{l-m}(1-Z)^{i-l}
.
\label{Xgen}
\end{align}

Vertices which have $k=1$ become active if they have a single active
neighbor. Thus a single seed vertex will activate an entire connected
cluster of such vertices. In particular, if there is a giant
connected cluster in the network (i.e. if $p \geq p_c$, the percolation
threshold), the introduction of a finite number of seed vertices into the
infinite network will (almost surely) activate the giant connected
component. In other words, we have behavior equivalent to ordinary
percolation. 
In particular, if we set $Q(0) + Q(1) = 1$ (and requiring that, if
$Q(0)\to 0$, the number of seed vertices remains sufficient to activate
the giant component of the network) then we
recover from Eqs. (\ref{Sgc_gen}) and (\ref{Xgen}) the well known
percolation equations \cite{dgm2008}. 

\section{Discussion}

In this paper we have extended the understanding of bootstrap
percolation to uncorrelated infinite random graphs with arbitrary
degree distribution, and studied the effects of damage to the network.
We have found that the phase diagram for the giant active component
with respect to damage to the network ($1-p$) and the fraction of initially
active vertices ($f$) has several interesting features. There are two
transitions observed. At the first the giant active component
appears continuously from zero, and at the second (always at a higher
initial activation fraction) there is a hybrid phase transition, where
the size of the giant active component has a discontinuity---a
`jump'---while also
having a singularity, as the size of the giant active component
approaches the transition from below as the square root of the
distance from the critical point. 
This singular behavior is due to avalanches in the activation process.
The sizes of avalanches of activation are determined by the size of
subcritical clusters---clusters of vertices whose number of active
neighbors is exactly one less than the activation threshold.
Everywhere but at the hybrid transition these subcritical clusters are
finite (though together occupying a finite fraction of the 
network), but as the transition is approached these clusters grow as
the reciprocal of the square-root of the distance from the transition.
We also observe a new special
critical point, at the level of damage at which the second transition
first appears. Here the height of the jump tends to zero, and the
scaling near the critical point is the cube-root of the distance from
the threshold. 
These results are valid for arbitrary degree
distributions, so long as they decay rapidly enough that the second
and third moments of the distribution are bounded.
Note that we could express our results not in terms of $f$ and $p$,
but of $f$ and any other convenient parameter, for example, the mean
degree $\langle q \rangle$ of a network. This allows one to apply our
conclusions to arbitrary uncorrelated networks.

Network inhomogeneity plays an important role.
When the second moment of the degree distribution is bounded but the
third moment is unbounded, the critical scaling near the appearance of
the giant active component is not simply linear but has higher order
scaling, depending on the degree distribution.
When the second moment is unbounded, for example in 
scale-free  
networks with degree distribution exponent $\gamma\leq 3$, 
both thresholds tend to $f=0$ and $p=0$ in the infinite size
limit. Thus the phase diagram is featureless, with a giant active
component (albeit sometimes very small) present for any finite
activation and any amount of damage to the network.
This result has important implications for real world networks. For
example, the network of neurons in the brain may have such a
scale-free organization \cite{Sporns2004} meaning that brain activity
may be able 
to be instigated with very small stimulus (even though such
networks, whilst large, are of course finite). 

In summary, we have obtained phase diagrams for the bootstrap
percolation problem in a wide range of complex networks. We have
described the properties and the nature of two distinct transitions in
this problem: the bootstrap percolation transition and the emergence
of a giant connected component (percolative cluster) of active vertices.

\begin{acknowledgments} 

This work was partially supported by the following projects PTDC: FIS/71551/2006, FIS/108476/2008, and SAU-NEU/103904/2008, and also by SOCIALNETS EU project. 

\end{acknowledgments} 

\bibliography{network}

\end{document}